\documentclass{usetex-v1}

\usepackage{algorithm}
\usepackage{algpseudocode}
\usepackage{amsmath}
\usepackage{array}
\usepackage{cite}
\usepackage{color}
\usepackage{graphicx}
\usepackage{multirow}
\usepackage{float}
\usepackage[labelformat=simple]{subcaption}
\usepackage{hyperref}
\usepackage[hyphenbreaks]{breakurl}
\usepackage{enumitem}

\renewcommand{\paragraph}[1]{{\smallskip\noindent\bf #1}}

\begin{document}

\title{Erasure Coding for Small Objects in In-Memory KV Storage}

\author{Matt M. T. Yiu, Helen H. W. Chan, and Patrick P. C. Lee\\
Department of Computer Science and Engineering, The Chinese University
of Hong Kong}

\maketitle


\begin{abstract}
We present MemEC, an erasure-coding-based in-memory key-value (KV) store that
achieves high availability and fast recovery while keeping low data redundancy
across storage servers.  MemEC is specifically designed for workloads
dominated by small objects.  By encoding objects in entirety, MemEC is shown
to incur 60\% less storage redundancy for small objects than existing
replication- and erasure-coding-based approaches.  It also supports graceful
transitions between decentralized requests in normal mode (i.e., no
failures) and coordinated requests in degraded mode (i.e., with failures).
We evaluate our MemEC prototype via testbed experiments under read-heavy and
update-heavy YCSB workloads.  We show that MemEC achieves high throughput
and low latency in both normal and degraded modes, and supports fast
transitions between the two modes. 
\end{abstract}

%
%
%


\section{Introduction}
\label{sec:intro}

Memory-centric storage systems have been proposed to keep primary data in
memory to enable scalable, low-latency data access.  They are often realized
as {\em in-memory key-value (KV) stores}, 
which organize data in memory as KV pairs (called {\em objects}) to
form structured storage.  
Enterprises have deployed in-memory KV stores for low-latency
operations in social networking, web search, and analytics (e.g., 
\cite{aniszczyk12,nishtala13,elasticache}).


Failures are prevalent in distributed storage \cite{ford10}.
To maintain data availability, most existing in-memory KV stores replicate
data copies across storage servers
\cite{nishtala13,repcached,aerospike,couchbase,redis,klein14}.
However, replication incurs high storage overhead and multiplies the
memory cost, which is still much higher than the cost of traditional
disk-based storage.  RAMCloud \cite{ousterhout15} keeps replicas in secondary
storage for persistence, yet accessing secondary storage for failure recovery
incurs high latency, especially for random I/Os (e.g., the typical hard disk
seek time is around 10ms).  

This motivates us to explore {\em erasure coding} for in-memory KV storage.
Erasure coding transforms data into fixed-size encoded chunks, such that the
original data can be recovered from a subset of encoded chunks.  It 
achieves the same fault tolerance with much
lower redundancy than replication \cite{weatherspoon02}.  Thus, we can
leverage erasure coding to keep minimum data redundancy {\em entirely} in
memory. Failure recovery can then be done by directly accessing other working
memory-based storage servers for encoded chunks, thereby maintaining low
access latency.   
A drawback of erasure coding is its high performance overhead in terms of
network bandwidth and disk I/O, especially in data updates and failure
recovery.  Thus, extensive studies focus on solving the performance issues of
erasure-coded storage (e.g., \cite{khan12,huang12,sathiamoorthy13}).

To deploy erasure coding in in-memory KV storage, we argue that two specific
challenges need to be addressed in addition to mitigating overheads in data
updates and failure recovery.
First, Facebook's field study \cite{atikoglu12} shows that real-life KV
storage workloads are dominated by {\em small objects}, whose keys and values
are of small sizes (e.g., ranging from few bytes to tens or hundreds of
bytes).  In particular, one workload has keys with either 16 or 21~bytes and
almost all values with 2~bytes only; another workload has up to 40\% of values
with only 2, 3, and 11~bytes \cite{atikoglu12}.  It is infeasible to perform
erasure coding directly on an extremely small object (e.g., 2~bytes),
as we need to first decompose the object into chunks before encoding
(\S\ref{sec:background}).  Second, to minimize access latency, in-memory KV
stores issue decentralized requests without centralized metadata lookups.
However, when failures happen, ongoing requests may need to be reverted or
replayed to avoid inconsistency.  In erasure coding, we also need to maintain
consistency across encoded chunks that are dependent on each other.
		 



We present MemEC, an erasure-coding-based in-memory KV store with access
performance, storage efficiency, and fault tolerance in mind.  MemEC works as
a {\em high-availability distributed cache} that supports low-latency data
access for read-intensive \cite{atikoglu12} or update-intensive
\cite{soundararajan10,chan14} workloads, and also enables fast recovery by
accessing data redundancy from other working in-memory storage servers
without the need of accessing secondary storage. 


MemEC supports small objects via a new {\em all-encoding} data model,
which encodes objects in entirety, including keys, values, and metadata, so as
to significantly reduce storage redundancy for fault tolerance.  We carefully
design the index structures for the all-encoding data model to reduce memory
footprints, such that all objects and index structures are kept entirely in
memory.  Our analysis shows that our all-encoding data model saves up to
60\% of redundancy over the replication- and erasure-coding-based approaches
used by existing in-memory KV stores.  
See \S\ref{sec:data_model}. 

MemEC allows decentralized requests for reads, writes, updates, and deletes in
normal mode (i.e., no failures), which is the common case in practice.  It
ensures graceful transitions between normal mode and degraded mode (i.e., with
failures), such that each request in degraded mode will be centrally
coordinated and redirected from a failed server to another working server.  
It also maintains both availability and consistency during transitions.  
See \S\ref{sec:overview} and \S\ref{sec:fault}. 

We implement a MemEC prototype that can be fully deployed on commodity
hardware in a cloud environment.
We conduct testbed experiments under read-heavy and update-heavy workloads
generated by YCSB \cite{cooper10}.  In normal mode, MemEC
achieves comparable performance to both Memcached \cite{memcached} and Redis
\cite{redis} (e.g., latencies are in the range of few milliseconds).  It also
efficiently operates in degraded mode, and completes transitions between
normal and degraded modes within milliseconds.  
See \S\ref{sec:implementation} and \S\ref{sec:evaluation}. 

The source code of our MemEC prototype is available at
{\bf https://github.com/mtyiu/memec}.



\section{Erasure Coding}
\label{sec:background}

Erasure coding is typically constructed by two configurable parameters $n$ and
$k$ (where $k < n$).  We treat data as a collection of fixed-size units called
{\em chunks}.  Every $k$ original chunks (called {\em data chunks}) are
encoded into $n-k$ additional equal-size coded chunks (called {\em parity
chunks}), such that the set of the $n$ data and parity chunks is called a 
{\em stripe}.  We consider erasure codes that are {\em maximum distance
separable (MDS)}, meaning that the encoding operations ensure that any $k$ of
the $n$ data and parity chunks of the same stripe can sufficiently decode the
original $k$ data chunks, while incurring the minimum storage redundancy
(i.e., $\frac{n}{k}$ times the original data size).  Reed-Solomon (RS) codes
\cite{reed60} are one well-known example of MDS erasure codes.


A storage system contains multiple stripes of data that are independently
encoded, and hence our analysis focuses on a single stripe.  The $n$ chunks of
each stripe are stored in $n$ distinct servers to tolerate server failures.
We call the server that stores a data (parity) chunk a {\em data (parity)
server}.  Since each stripe is stored in a different set of servers, a server
may act as either a data server or a parity server for different
stripes; in other words, the naming of a data or parity server is logical. 


Each update to a data chunk will trigger an update to every parity chunk of the
same stripe.  To mitigate the network overhead of parity updates, we 
leverage the {\em linearity} property of erasure coding without 
transferring any existing parity chunk for the update.  For example, RS codes
encode data chunks by linear combinations based on Galois Field
arithmetic.  Specifically, a parity chunk (denoted by $P$) is encoded by
$k$ data chunks (denoted by $D_1$, $D_2$, $\cdots$, $D_k$) via a linear
combination as $P = \sum_{i=1}^k \gamma_i D_i$ for some encoding coefficients
$\gamma_i$'s ($1\le i\le k$).  Suppose that a data chunk $D_i$ is now updated
to $D_i'$.  We can compute the new parity chunk as $P' = P + \gamma_i (D_i' -
D_i)$, where $(D_i' - D_i)$ is called the {\em data delta}.  Thus, when a data
server applies an update to a chunk, it sends the data delta directly to
each parity server, which can then use the data delta to compute the new
parity chunk.  Previous studies have exploited the linearity property for
parity updates in RAID \cite{chen94} and distributed storage
\cite{aguilera05,chan14,zhang16}.  


Erasure coding generally requires the chunk size be large enough. For example,
RS codes require $n\!\le\!2^w\!-\!1$, where $w$ is the
number of bits of a chunk \cite{plank09}.  To deploy RS codes, it is typical
to set $w$ as a multiple of 8 to align with machine word boundaries; in other
words, a chunk needs to be at least one byte long.  For an extremely small
object (e.g., 2 bytes long \cite{atikoglu12}), dividing it into $k$ (e.g., $k
> 2$) data chunks for encoding can be infeasible. 


\section{Data Model}
\label{sec:data_model}

In this section, we propose a data model for small objects in in-memory KV
stores.  Each object in a KV store comprises three fields: a key, a value, and
metadata, such that both the key and value have variable sizes, while we
assume that the metadata has a fixed size for simplicity.  The key is a unique
identifier for an object; the value holds the actual content of the
object; the metadata holds the object attributes such as the key size, the
value size, creation/modification timestamps, etc.  Also, the KV store
has an {\em index structure} that holds the references to all objects for
object retrievals.  



\subsection{Existing Data Models}
\label{subsec:existing_data_models}

We revisit two existing data models that achieve fault tolerance via
redundancy, namely all-replication and hybrid-encoding.  Both data models have
been realized in existing in-memory KV stores.

{\bf All-replication} stores multiple replicas for each object,
including the key, value, metadata, and reference to the object.
It has been used in Facebook's Memcache \cite{nishtala13}, DXRAM
\cite{klein14}, RAMCloud \cite{ousterhout15}, and various open-source
in-memory KV stores (e.g., \cite{repcached,aerospike,couchbase,redis}).
If there is a failed object, the KV store retrieves its replica from another
working server.  Most KV stores directly store replicas in memory.  To reduce
memory footprints, RAMCloud stores replicas in secondary storage, but this
incurs expensive I/Os when there are intensive random reads to failed objects. 

{\bf Hybrid-encoding} combines erasure coding and replication.
It applies erasure coding to the value only, while applying
replication for the key, metadata, and reference to the object. 
Its rationale is that if the value size is significantly large, applying
erasure coding to values only suffices to reduce storage redundancy. 
Specifically, hybrid-encoding performs $(n,k)$ erasure coding across object
values, such that the values of multiple objects are combined to form a data
chunk, and the $k$ data chunks in the same stripe are encoded to $n-k$ parity
chunks.  
For each object, hybrid-encoding replicates the key, metadata, and any index
information across $n-k+1$ servers, including the data server that stores the
object (which is embedded in the data chunk) and the $n-k$ parity servers.
Thus, every object can tolerate any $n-k$ server failures.  
Hybrid-encoding has been adopted by existing KV stores, such as LH*RS
\cite{litwin05}, Atlas \cite{lai15}, and the recently proposed in-memory KV
store Cocytus \cite{zhang16}.  In particular, Cocytus focuses on the value
size of at least 1KB in evaluation \cite{zhang16}.  The drawback of
hybrid-encoding is that if the value size is very small, it still incurs
significant storage overhead due to the replication of keys and metadata.

\subsection{Our Data Model}
\label{subsec:our_data_model}

We now propose a new data model called {\bf all-encoding}, whose idea is to
apply erasure coding to objects {\em in entirety} (including keys, values, and
metadata) without replication.  In addition, we carefully design our index
structure to limit its storage overhead.

\begin{figure}[!t]
\centering
\includegraphics[width=3in]{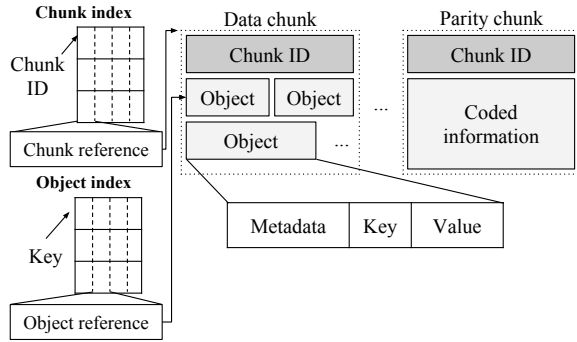}
\vspace{-4pt}
\caption{All-encoding data model. Each server keeps a local copy of both
object and chunk indexes, while the chunks are distributed across servers.}
\vspace{-6pt}
\label{fig:data_model}
\end{figure}

\paragraph{Data organization:} Figure~\ref{fig:data_model} shows the data
organization of the all-encoding model.  We organize data as fixed-size
chunks, which form the units of erasure coding.  A larger chunk size provides
better storage efficiency by mitigating the storage overhead due to indexing
(see details below), yet it also incurs higher computational overhead in
encoding/decoding.  Our current implementation chooses 4KB as the default
chunk size to balance the trade-off.  For each chunk, we prepend a unique
fixed-size {\em chunk ID} (8~bytes in our case) for chunk identification in a
server.

All-encoding performs erasure coding across objects in entirety.  
Suppose that every object is small, such that its total size is
smaller than the chunk size (i.e., the object can fit into a single chunk).
Thus, each data chunk can contain multiple objects, each of which starts with
metadata, followed by a key and a 
value.  We append new objects to a data chunk until the chunk size limit is
reached, and {\em seal} the data chunk so that no new objects can be further
added.  Sealed data chunks will later be encoded to form parity chunks
belonging to the same stripe. 

For indexing, we introduce two hash tables: the {\em chunk index}, which maps
a chunk ID to a chunk reference, and the {\em object index}, which maps a key
to an object reference.  We emphasize that we do not need to keep
redundancy for both indexes for fault tolerance; instead, each 
server only keeps a local copy of both indexes.  The reason is that
when a server fails, we can reconstruct the failed chunks from other available 
servers by erasure coding, and rebuild both indexes by reinserting the
references of the reconstructed objects and chunks into the object and chunk
indexes, respectively.  In other words, the redundancy of both indexes is
implicitly included in chunk-level redundancy.

To improve storage efficiency of both indexes, we apply {\em cuckoo hashing}
\cite{pagh04}, which has constant lookup and amortized update times.
Its idea is to use two hash functions to map a key to two possible locations
(called {\em buckets}).  Each inserted key is stored in either free bucket; if
no free bucket is available, we relocate any existing key to make room for the
inserted key.  We set both indexes to be 4-way set-associative (i.e., each
bucket contains four slots, each storing a reference), and it is shown that
the space utilization can reach over 90\% \cite{fan13}.
Currently, we do not consider range queries, which can be supported by
tree-based indexes \cite{zhang15}.   

\paragraph{Chunk ID:}  To support updates and recovery, we need to locate the
data and parity chunks of the same stripe.  We leverage the chunk ID for this
purpose.  We decompose a chunk ID into three fields: 
(i) {\em stripe list ID}, which identifies the set of $n$ data and parity
servers that stores the stripe associated with the chunk 
(\S\ref{subsec:decentralized}), (ii) {\em stripe ID}, which identifies a
stripe in the storage system, and (iii) {\em chunk position}, which numbers
the position of the chunk in a stripe from $0$ to $n-1$.  In particular, the
data and parity chunks of the same stripe will have the same stripe list ID
and same stripe ID. 
To obtain a stripe ID, each server maintains a local counter (initialized from
zero) for each stripe list.  Each time when a chunk is sealed, the data server
sets the stripe ID as the current counter value, followed by incrementing the
counter value by one.  Each parity server encodes the data chunks of the same
stripe into a parity chunk, which inherits the stripe list ID and stripe ID
from the data chunks. 

We use the chunk ID and both object and chunk indexes to locate an object.
We align all fixed-size chunks (including the 8-byte chunk ID
and 4KB chunk content) in the address space of each server.   To update an
object, we locate the object in the data server via the object index, and then
obtain the corresponding chunk ID that is stored at the head of the chunk.
Using the chunk ID, we identify each corresponding parity server, in which we
can locate the parity chunk of the same stripe via the chunk index.

We also need to maintain the mappings between each key and chunk ID, so that
when we reconstruct a failed object during a server failure, we can identify
its associated chunk ID and retrieve the chunks of the same stripe in other
servers for reconstruction.  Nevertheless, the mappings can be stored in
secondary storage, since they are only needed for recovery from failures.
Thus, the mappings do not add overhead to in-memory storage.  We elaborate how
we maintain key-to-chunk mappings in \S\ref{subsec:fault_backup}.


\paragraph{Handling large objects:}
Some workloads can have objects with large value sizes (e.g., 1MB
for the ETC workload in Facebook \cite{atikoglu12}).   We can extend our data
model to store a large object that cannot fit into a chunk, assuming that the
key size remains small.  We divide an object into {\em fragments}, each of
which has the same size as a chunk (except the last fragment whose size may be
less than the chunk size).  We store each fragment as an object, and include
an offset field in the object's metadata to specify the position of each
fragment.  We encode the fragments, together with other chunks in the same
stripe, via erasure coding.  Note that all fragments keep both key and
metadata; in particular, the same key is replicated across fragments.
However, since the value size is now large, the storage overhead incurred by
the key and metadata becomes low. 

\subsection{Analysis}
\label{subsec:data_model_analysis}

We analyze the redundancy overhead of the above data models.  Our
analysis makes the following assumptions.  We only focus on the redundancy
incurred for fault tolerance.
Also, for all-replication and hybrid-encoding, we
only include the index overhead for locating objects that are stored locally,
while excluding their index overhead for maintaining the correlation between
the original data and its redundancy since such implementation varies across 
systems.  Thus, our analysis gives {\em underestimates} for their redundancy
overhead.  On the other hand, for all-encoding, our analysis includes the
index overhead due to chunk IDs and both object and chunk indexes, while 
excluding the key-to-chunk mappings as the latter can be stored in
secondary storage and only used when failures happen
(\S\ref{subsec:our_data_model}).  For simplicity, our analysis assumes fixed 
key and value sizes.  

\begin{figure}[t]
\centering
\begin{tabular}{c@{\ }c}
\includegraphics[width=1.55in]{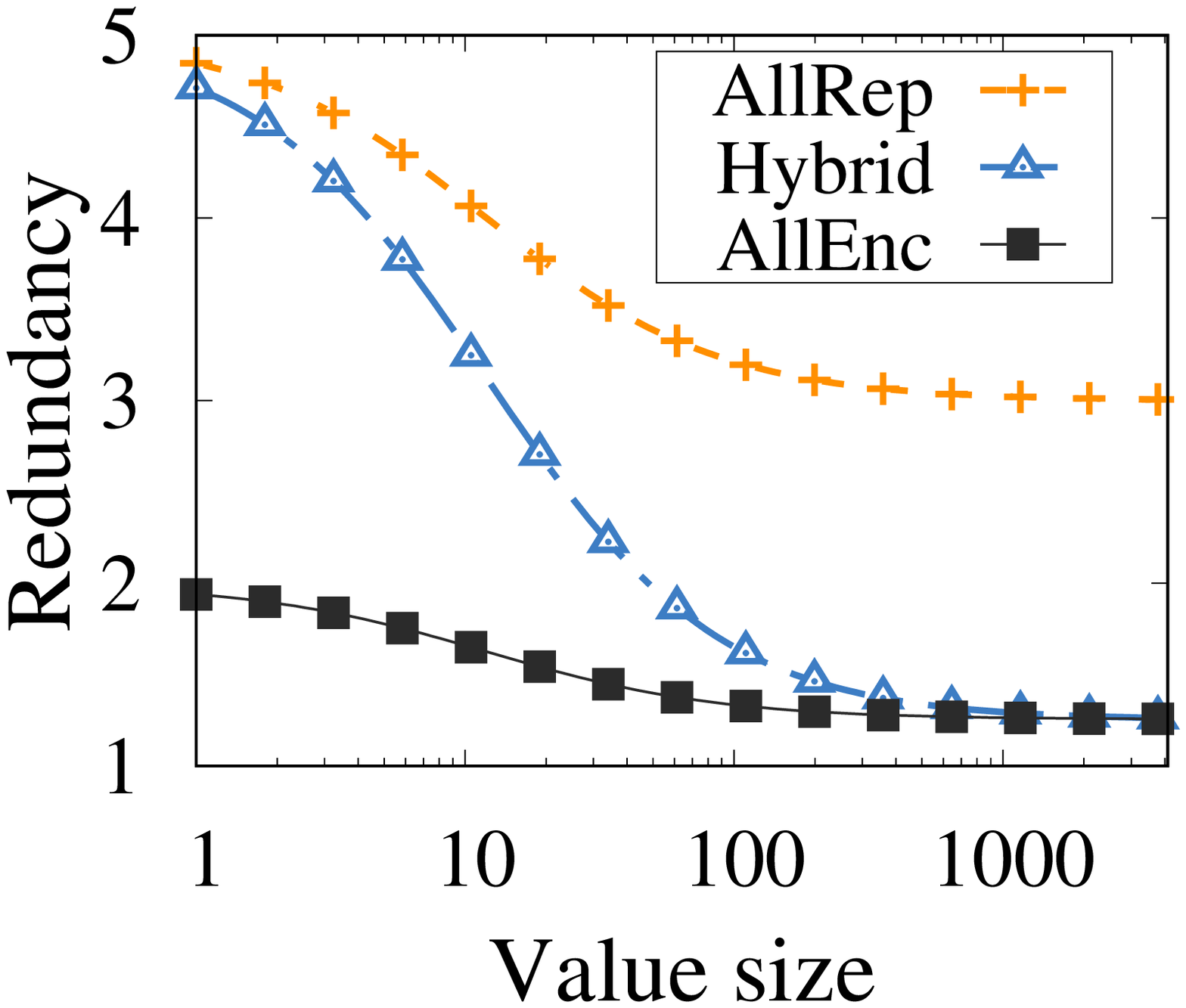} &
\includegraphics[width=1.55in]{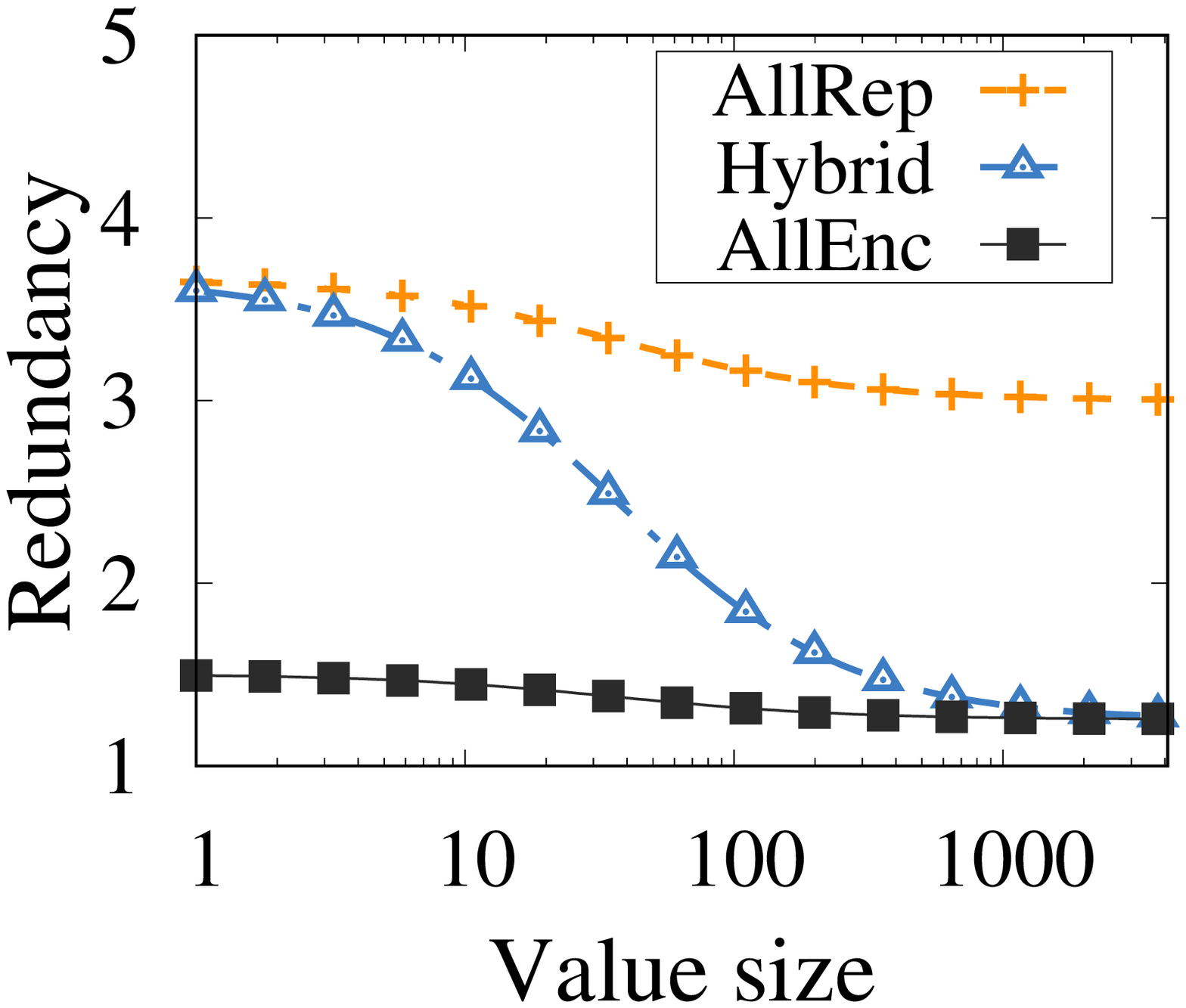} \\
\mbox{\small (a) $K\!=\!8$, $(n,k)\!=\!(10,8)$} &
\mbox{\small (b) $K\!=\!32$, $(n,k)\!=\!(10,8)$} \\
\includegraphics[width=1.55in]{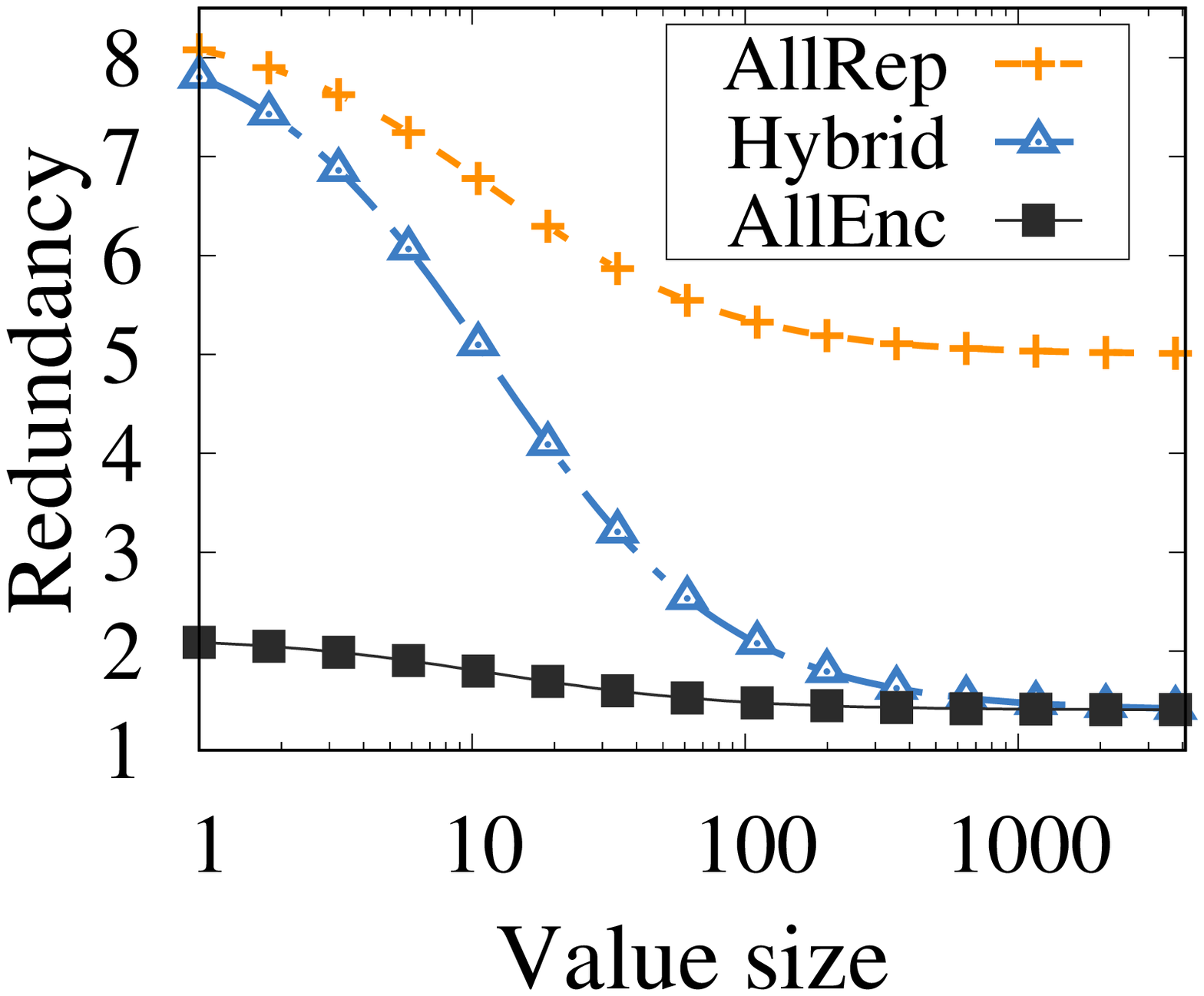} &
\includegraphics[width=1.55in]{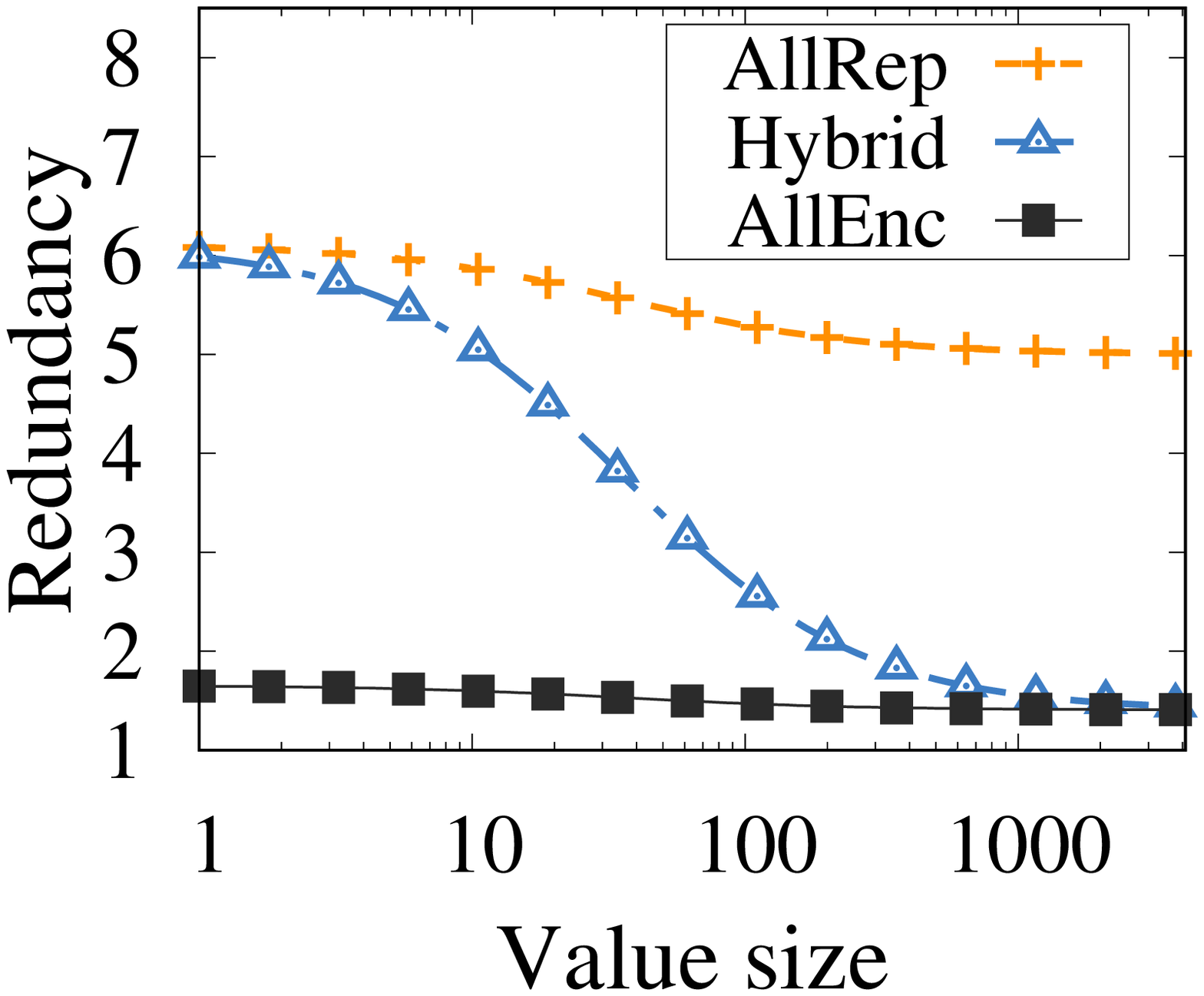} \\
\mbox{\small (c) $K\!=\!8$, $(n,k)\!=\!(14,10)$} &
\mbox{\small (d) $K\!=\!32$, $(n,k)\!=\!(14,10)$}
\end{tabular}
\vspace{-3pt}
\caption{Redundancy of all-replication (AllRep), hybrid-encoding (Hybrid), and
all-encoding (AllEnc).}
\label{fig:redundancy}
\vspace{-6pt}
\end{figure}

Let $K$, $V$, $M$, and $R$ be the key size, value size, metadata size, and the
size of a reference, respectively.  We quantify the redundancy of a data model
as the ratio of the actual storage size of an object with redundancy to the
original size of the object without redundancy incurred by fault tolerance
(i.e., $K+V+M$).  

\paragraph{All-replication:} To tolerate $n-k$ failures, we make
$n-k+1$ copies for each object.  The redundancy is:
\vspace{-3pt}
\[
\frac{(n-k+1)(K + V + M + R)}{K + V + M}. 
\]
\vspace{-6pt}

\paragraph{Hybrid-encoding:}  To tolerate $n-k$ failures, we replicate the
key, metadata, and reference of each object with total size $(n-k+1)(K+M+R)$
and encode the value by erasure coding with total size $\frac{nV}{k}$.
The redundancy is:
\vspace{-3pt}
\[
\frac{(n-k+1)(K + M + R) + \frac{nV}{k}}{K + V + M}. 
\]
\vspace{-6pt}

\paragraph{All-encoding:}  We encode the key, value, and metadata of each
object in a data server and $n-k$ parity servers with total size $\frac{n(K +
V + M)}{k}$.  The data server also stores a reference to the object in its
object index with total size $\frac{R}{O}$, where $O$ is the occupancy of the
object index due to cuckoo hashing.  Each chunk also has a chunk ID, which we
assume has size $I$, and a reference in the chunk index with total size
$\frac{R}{O}$ (also due to cuckoo hashing).  Let $C$ be the chunk size, so the
$n$ chunks of a stripe store $\frac{kC}{K + V + M}$ objects on average.  Thus,
the average storage size of the chunk ID and chunk reference per object is
$\frac{n(I + R / O)}{kC / (K + V + M)}$.  The redundancy is:
\vspace{-3pt}
\[
\frac{\frac{n(K + V + M)}{k} + \frac{R}{O} + \frac{n(I + R / O)}{kC / (K + V + M)}}{K + V +
M}.
\]
\vspace{-6pt}

\paragraph{Numerical results:}  We analyze the redundancy
of all data models for different $K$, $V$, and $(n,k)$.  
We fix $M\!=\!4$~bytes (assuming that the metadata only holds a 1-byte key size
and 3-byte value size), $R\!=\!8$~bytes, $C\!=\!4$KB, $I\!=\!8$~bytes (for a
chunk ID in all-encoding), and $O\!=\!0.9$ (\S\ref{subsec:our_data_model}). 

Figure~\ref{fig:redundancy} presents the results for $K\!=\!8$ and $32$~bytes,
$(n,k)\!=\!(10,8)$ and $(14,10)$ (note that $(14,10)$ is used by Facebook's
erasure coding \cite{muralidhar14}), and different values of $V$ (in bytes).
All-encoding significantly reduces redundancy especially for small key and
value sizes.  For example, when $K\!=\!8$, $V\!\le\!10$, and
$(n,k)\!=\!(10,8)$ (Figure~\ref{fig:redundancy}(a)), all-replication and
hybrid-encoding achieve 4.1--4.8$\times$ and 3.3--4.7$\times$ redundancy,
respectively, while all-encoding reduces the redundancy to 1.7--1.9$\times$
(up to 60.0\% and 58.9\% reduction, respectively).  We make similar findings
for $K\!=\!32$ and $(n,k)\!=\!(14,10)$.   Both hybrid-encoding and
all-encoding are approaching $\frac{n}{k}$ of redundancy as $V$ increases, but
all-encoding is clearly faster.   For example, when $(n,k)\!=\!(10,8)$ and 
$K\!=\!8$, the redundancy of all-encoding drops to below 1.3 (4\% over
$\frac{n}{k}\!=\!1.25$) when $V\!\ge\!180$, while hybrid-encoding achieves the
same value when $V\ge890$.

\section{MemEC Overview}
\label{sec:overview}

\subsection{System Architecture}
\label{subsec:arch}

\begin{figure}[t]
\centering
\includegraphics[width=2.8in]{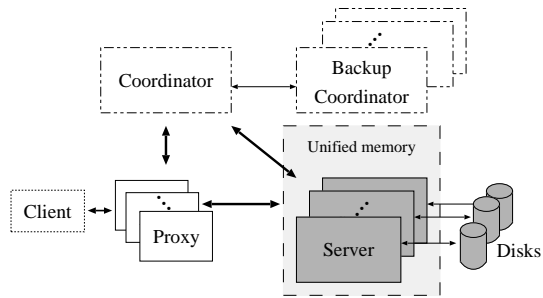}
\vspace{-3pt}
\caption{MemEC architecture.}
\vspace{-6pt}
\label{fig:arch}
\end{figure}

MemEC is an in-memory KV store that manages the storage of objects with the
all-encoding data model.
Figure~\ref{fig:arch} shows the MemEC architecture, which comprises three
types of nodes: multiple {\em servers}, multiple {\em proxies}, and a 
{\em coordinator}.  Each server represents a storage node and allocates a
memory region for storage.  It is attached to persistent secondary storage
(e.g., local disks or networked file systems).  We aggregate the
memories of all servers to a unified in-memory storage pool.  Each proxy
is a front-end interface for client applications to access objects in
servers.  The coordinator manages the architecture, detects failures (e.g., by
periodic heartbeats), and coordinates I/O requests in the presence of
failures.  Our current prototype uses a single coordinator for simplicity, 
yet we can synchronize its state to multiple backup coordinators or external
reliable storage for fault tolerance.  We can also deploy a distributed
coordinator service via Zookeeper \cite{hunt10} for improved performance. 

MemEC maintains data availability due to server failures, in which in-memory
storage becomes unavailable.  When there is no server failure, MemEC operates
in {\em normal mode}; when a server fails, the coordinator notifies all
working proxies and servers to switch to {\em degraded mode} and
coordinates the reconstruction of any failed data.  In normal mode, MemEC
distributes objects from each proxy across servers in a decentralized manner
without involving the coordinator, as in existing in-memory KV stores
\cite{fitzpatrick04,ousterhout15}.
In contrast, in degraded mode, the coordinator is included into the I/O path
and responsible for redirecting read/write requests away from any failed
server.  Note that if a request does not involve any failed server, it is
still executed in a decentralized manner.   Our design choice builds on the
fact that the common case in practice is that MemEC runs in normal mode and
uses decentralized requests for most of the time, while the coordinator is
only involved in the I/O path when failures happen. 
In \S\ref{sec:fault}, we elaborate our failure model, and state additional
design assumptions for fault tolerance.

MemEC supports linearizability. It keeps each object in only one working
server in both normal and degraded modes.  All servers process requests in a
first-in-first-out manner, so that each read to an object always returns its
value of the last write.


\subsection{Stripe Management}
\label{subsec:decentralized}


In normal mode, MemEC distributes objects across servers in a decentralized
manner.  During bootstrap, MemEC generates a pre-configured number (denoted by
$c$) of {\em stripe lists}, each of which defines a set of $k$ data servers
and $n-k$ parity servers of an erasure-coded stripe.  Then a proxy maps an
object to a server via two-stage hashing: it first hashes the object's key to
one of the stripe lists, followed by hashing the object's key to one of the
data servers in the stripe list.

To generate the stripe lists, one concern is that in write-intensive
workloads, each write to a data server triggers a write to each parity server
of the same stripe, so a parity server receives $k$ times more write load than
a data server (assuming that all data servers receive the same amount of write
loads).  Thus, we consider an algorithm that iteratively generates $c$ stripe
lists, with an objective of balancing the write loads across all servers.
Specifically, we assign a load variable, initialized to be zero, to each
server in MemEC.  In each iteration $i$ (where $1\le i\le c$), we select $n-k$
servers with the smallest loads as the parity servers, followed by $k$ servers
with the next smallest loads as data servers (we break any tie by choosing
servers with smaller IDs).  We form the $i$-th stripe list with the $n$
selected servers.  We increment the load of each data server by $1$ and that
of each parity server by $k$.  Finally, the algorithm outputs $c$ stripe
lists, which are installed into all proxies and servers.  Note that the
algorithm is executed only once during system startup, and runs in polynomial
time with respect to the number of servers and the number of stripe lists.
Thus, it has limited overhead to overall system performance.

\subsection{Basic Requests}
\label{subsec:basic}

MemEC realizes four basic requests of practical KV stores:
\texttt{SET}, \texttt{GET}, \texttt{UPDATE}, and \texttt{DELETE}.
Here, we focus on small objects, and handle large objects as described in 
\S\ref{subsec:our_data_model}.  Also, we focus on normal mode, and address
degraded mode in \S\ref{sec:fault}.

\paragraph{SET} inserts a new object into storage. A proxy first selects one
data server and $n-k$ parity servers as described in
\S\ref{subsec:decentralized}.  It sends the object to these servers in
parallel, each of which returns an acknowledgement upon receiving the object.
The data server appends the object to a data chunk that is not yet full 
(i.e., {\em unsealed}), while each parity server keeps the object in a
temporary buffer as a replica.  The data server also adds the object
reference to its own object index.  Note that the data and parity servers do
not need to communicate with each other upon a \texttt{SET} request.  When the
data chunk is full, it is {\em sealed} (\S\ref{subsec:our_data_model}). 
The data server then sends the keys of
the objects in the sealed data chunk to all parity servers.  Upon receiving
the keys, each parity server rebuilds the data chunk from the objects in its
temporary buffer (as the objects may not arrive in the same order), computes
the data delta and updates its parity chunk (\S\ref{sec:background}), and
finally discards the objects from its temporary buffer.

In essence, our \texttt{SET} operation performs $(n\!-\!k\!+\!1)$-way
replication until a chunk is sealed, and then applies erasure coding to reclaim
the space of replicas.  The transition from replication to erasure coding only
involves the transmissions of object keys, and hence poses limited bandwidth
overhead.  

Each server is initialized with a pre-configured number of chunks based on the
available storage capacity, and maintains a list of currently unsealed data
chunks during execution.  When it receives a new object via a \texttt{SET}
request and is selected as the data server for storing the object, it appends
the object to the unsealed data chunk that has the minimum remaining free
space sufficient for holding the object, so as to fill up the unsealed data
chunk as soon as possible.  If all unsealed data chunks cannot store the new
object, the server will pick the unsealed data chunk with the least free space
to seal, so as to make a free unsealed data chunk available.  Since we fix the
number of unsealed data chunks, it also ensures that the number of replicas in
the temporary buffer (for unsealed data chunks) of each parity server is small,
thereby keeping the storage redundancy low. 

\paragraph{GET} retrieves an object from storage. A proxy uses the key to
determine the data server that stores the object, and requests the data server
for the object's value.

\paragraph{UPDATE} modifies an existing object with a new value. A proxy
first identifies the data server that stores the object, and sends the new
value to the data server.  The data server then updates the object's value and
forwards the data delta to all parity servers of the same stripe; the set of
parity servers can be determined by the chunk ID
(\S\ref{subsec:our_data_model}).
Each parity server either updates the corresponding parity chunk if the object
is in a sealed data chunk, or the replica in the temporary buffer if the
object is in an unsealed data chunk.  After all parity servers finish the
update, the data server returns an acknowledgement to the proxy.

Here, we allow different value sizes across objects, yet we assume that the
value size of an object remains unchanged across updates. This keeps the size
of a sealed data chunk unchanged, thereby simplifying storage management. How
to adjust the object's value size on-the-fly is a future work.

\paragraph{DELETE} removes an object from storage. A proxy requests the data
server that stores the object to mark the object as deleted.  If the object is
in a sealed data chunk, the data server sends the data delta to all parity
servers of the same stripe as in \texttt{UPDATE}, by treating the new object's
value as zero.  Each parity server updates the parity chunk accordingly.
Otherwise, the data server notifies each parity server to remove its replica
from the temporary buffer.  The data server returns an acknowledgement to the
proxy after the data server and all parity servers complete the deletion.  We
can later reclaim the space of the deleted objects.
\section{Fault Tolerance}
\label{sec:fault}

In this section, we describe how MemEC achieves fault tolerance, and in
particular, ensures graceful transitions between normal and degraded modes.

\subsection{Failure Model}

Our failure model focuses on server failures that make data unavailable.
Server failures can be {\em transient} or {\em permanent}.  Transient failures
are temporary and do not incur actual data loss, such as due to server
overloads and network congestion; in contrast, permanent failures incur actual
data loss.  When transient failures occur, MemEC reconstructs the unavailable
data from other in-memory working servers for fast recovery; in contrast, when
permanent failures occur, MemEC can either recover lost data from other
working servers or from secondary storage (Figure~\ref{fig:arch}).  We require
that the coordinator be responsible for detecting server failures, say by
periodic heartbeats (\S\ref{subsec:arch}).  In this work, we focus on how
MemEC reacts when transient server failures happen and performs fast recovery. 




We do not consider proxy failures when MemEC is in degraded mode
(i.e., server failures exist).  In this case, we assume that MemEC
reconstructs all available data from secondary storage.  Nevertheless, our
evaluation shows that resolving inconsistencies is fast (less than 700ms), so
it is relatively rare for this special case to happen.  On the other hand, when
there is no server failure, each proxy only acts as an interface for clients
to access servers.  If a proxy fails, clients can access objects in servers
through a different proxy.  We also do not consider coordinator failures, as
we can keep backups for the coordinator's states (\S\ref{subsec:arch}).

\subsection{Server States}
\label{subsec:fault_states}

To maintain availability and consistency, MemEC requires that all
requests that involve a failed server (called {\em degraded requests}) be
managed by the coordinator, which redirects a request from any failed server
to a different working server (called {\em redirected server}).  However,
making a graceful transition from normal mode to degraded mode is non-trivial.
Specifically, there may be ongoing requests that {\em inconsistently} update
the data and parity chunks of a stripe.  For example, in an \texttt{UPDATE}
request, a data server sends a data delta to all parity servers of the same
stripe (\S\ref{subsec:basic}).  If the data server fails, we cannot guarantee
that all parity servers receive the data delta and update their parity chunks,
since the failed data server is the only entity that keeps track of the parity
updates.  In this case, MemEC needs to resolve any potential inconsistency
caused by the incomplete request.

\begin{figure}[t]
	\centering
	\includegraphics[width=2.8in]{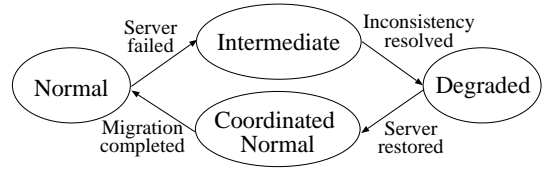}
	\vspace{-4pt}
	\caption{State diagram for failure handling.}
	\vspace{-6pt}
	\label{fig:states}
\end{figure}

To address the above challenges, in MemEC, the coordinator maintains
{\em server states} to explicitly specify the operational status of each
server.  Figure~\ref{fig:states} shows the server state diagram, which
specifies four states:
\begin{itemize}[leftmargin=*] \itemsep=0pt \parskip=0pt
\item
{\bf Normal state}:  It is the initial state of all servers, in which all requests are decentralized (\S\ref{subsec:basic}).
\item
{\bf Intermediate state}:  It is the state when a server fails while data
inconsistency is not yet resolved.
\item
{\bf Degraded state}: After data inconsistency is resolved, all proxies issue
degraded requests in collaboration with the coordinator.
\item
{\bf Coordinated normal state}:
When the failed server is restored, the coordinator instructs the redirected
server to migrate any changes made in the degraded state to the restored
server.  All proxies still issue requests through the
coordinator before the migration is completed.
\end{itemize}

Based on the current state of a server, a proxy decides whether decentralized
or degraded requests should be issued to the server; for degraded requests,
other working servers are also involved to reconstruct failed data.  Thus, we
require that {\em all proxies and working servers share the same view of
the server states}.  To do this, we need a two-phase protocol when the
coordinator detects a server failure: the coordinator first notifies all
proxies to complete processing requests in normal mode; after all proxies have
confirmed that they have completed all normal requests, the coordinator
notifies all proxies to start processing requests in degraded mode.  We
leverage {\em atomic broadcast} to realize the two-phase protocol, so that the
notifications from the coordinator in both phases are reliably delivered
to all proxies. 


We emphasize that MemEC only uses atomic broadcast to reliably notify
all proxies about state transitions in the presence of failures, but not for
handling normal and degraded requests.  Also, when a server fails, we do not
require all proxies to transit from normal mode to degraded mode atomically
(even they have all received the atomic broadcast message); instead, the
failed server switches to the intermediate state, in which the coordinator
notifies all proxies via atomic broadcast that they no longer need to issue
decentralized requests to the failed server.  After all proxies resolve
inconsistencies of incomplete requests (\S\ref{subsec:fault_backup}) and
notify the coordinator, the coordinator will issue another atomic broadcast to
notify all proxies that the server is safe to switch to the degraded state, in
which all requests are now handled through the coordinator.  We elaborate the
details in the following discussion.  Our evaluation shows that the state
transition overhead is small (\S\ref{sec:evaluation}).



\subsection{Backups}
\label{subsec:fault_backup}

MemEC resolves inconsistency when a failed server is in the intermediate
state.  To achieve this, it makes backups for ongoing requests, so as to
revert inconsistent changes or replay incomplete requests.  We emphasize that
such backups are only temporarily kept, and will be cleared when data
consistency is achieved.  MemEC keeps three types of backups, as explained
below.


Each proxy makes backups for unacknowledged requests, so that incomplete
requests due to the failed server can be replayed later as degraded requests.
Specifically, a proxy locally buffers each request that is sent to a server,
and clears the buffered request upon receiving an 
acknowledgment. 

Each proxy also makes backups for the mappings between keys and chunk IDs
generated by servers to facilitate the reconstruction of failed objects.  We
require that each data server periodically checkpoints the mappings between
keys and chunk IDs to the coordinator for recovery
(\S\ref{subsec:our_data_model}).  
When each data server returns an acknowledgement to a proxy request, it also
piggybacks the mapping between the object's key and the chunk ID associated
with the request.  The proxy then buffers the mapping.  When a server issues a
new checkpoint, it acknowledges each proxy to clear its buffered mappings.
During a server failure, each proxy sends its buffered mappings to the
coordinator for reconstructing the mappings from the latest checkpoint.


Furthermore, each parity server makes backups for the data deltas for both
\texttt{UPDATE} and \texttt{DELETE} requests, so as to revert any changes on
parity chunks made by incomplete (unacknowledged) requests.  We adopt an idea
similar to that in LH*RS \cite{litwin05}, in which we let each parity server
store the received data deltas in a temporary buffer for possible rollbacks.
When a data server fails, the proxy that initiates an incomplete
\texttt{UPDATE} or \texttt{DELETE} request receives the updated server state
from the coordinator (\S\ref{subsec:fault_states}).  It notifies the parity
servers to revert the changes on parity chunks for the incomplete request
using the buffered data deltas.
To facilitate the removal of buffered data deltas, when a proxy issues an
\texttt{UPDATE} or \texttt{DELETE} request, it also attaches a {\em local
sequence number} (i.e., independent of other proxies) for the latest
acknowledged request, so a parity server can decide if any buffered data delta
can be removed.



\subsection{Degraded Requests}
\label{subsec:fault_degraded}

When a failed server switches to the degraded state (i.e., after data
inconsistency is resolved), the coordinator redirects all degraded requests
from the failed server to a redirected server.  The following explains how
each type of degraded requests works.

\paragraph{Degraded SET:}  Recall that a \texttt{SET} request sends an object
to a data server and $n-k$ parity servers (\S\ref{subsec:basic}).
If a proxy finds that a \texttt{SET} request for
a new object is mapped to a failed data or parity server, it
issues a degraded \texttt{SET} through the coordinator.  The coordinator first
identifies the stripe list for the object.  It then finds a working server in
the stripe list as the redirected server, and instructs the proxy to write
the object to the redirected server.  Upon receiving the object, the
redirected server stores the object in a temporary buffer. Later when the failed server
is restored, the redirected server migrates the object to the restored server.



\paragraph{Degraded GET:}  If a proxy issues a \texttt{GET} request for an
object in a failed data server, the coordinator first maps the object to its
stripe list and selects a working server in the stripe list as the redirected
server (which can be the same as in degraded \texttt{SET}).  The proxy then
asks the redirected server for the object.   If the object
is stored in an unsealed data chunk, the redirected server simply retrieves
its replica from a working parity server; if the object is stored in a sealed
data chunk, the redirected server reconstructs the whole data chunk via
erasure coding to retrieve the failed object; if the object is written via a
degraded \texttt{SET}, the redirected server retrieves the object from
its temporary buffer.

Here, the redirected server performs {\em on-demand} reconstruction for the
data that is requested by the proxy, instead of reconstructing all data
stored in the failed server.  This avoids unnecessary decoding overhead,
especially in the presence of transient failures in which data is only
temporarily unavailable.
Moreover, the redirected server reconstructs a failed
object at the granularity of chunks (i.e., the whole chunk containing the
failed object will be reconstructed).  If subsequent \texttt{GET} requests
access the same object or other objects within the reconstructed chunk, no
extra reconstruction is needed.




\paragraph{Degraded UPDATE and DELETE:}  If a proxy issues an \texttt{UPDATE}
or \texttt{DELETE} to an object whose associated stripe list contains a failed
server, the request becomes a degraded request.  In this case, the coordinator
first identifies a working server in the stripe list as the redirected server,
which first reconstructs the failed chunk (as in degraded \texttt{GET}) before
the object is modified or deleted.  Note that we reconstruct the failed chunk
even though the object is available in a working server (belonging
to the same stripe list as the failed server).  The reason is that each update
or delete to an object also triggers updates to all parity chunks.  We ensure
that the failed chunk is reconstructed before its corresponding parity chunks
are updated, thereby eliminating the need of resolving any inconsistency.
After the failed chunk is reconstructed, the data server (which may be the
redirected server if the original data server fails) handles the request as
in \S\ref{subsec:basic}.  Note that if the failed server is a parity server,
the data server sends the data delta to the redirected server.

\subsection{After Failures}
\label{subsec:fault_post_recovery}


When a failed server is back to normal, it is in the coordinated normal state.
All redirected servers migrate any reconstructed objects back to the restored
server, while new
proxy requests are still managed by the coordinator.  Specifically, if a
proxy accesses an object that has been redirected before and not yet
migrated, the coordinator instructs the proxy to access the redirected
server; if the redirected object has been migrated, the coordinator instructs
the proxy to send the request to the restored server.  When all data
migration is completed, the restored server switches to the normal state, and
proxies can issue decentralized requests without involving the coordinator.





\section{Implementation}
\label{sec:implementation}

MemEC is deployable on commodity hardware and operating systems.  We have
implemented a MemEC prototype in C++ on Linux in about 34K lines of code.  We
use the Spread toolkit \cite{spread14} to implement atomic broadcast for
server state synchronization (\S\ref{subsec:fault_states}), and use Intel's
Storage Acceleration Library \cite{isal} to realize erasure coding. 

MemEC is optimized for high network performance.  It maintains persistent TCP
connections between the coordinator, proxies, and servers to save connection
establishment overhead.  It also uses non-blocking socket I/O
based on {\tt epoll} \cite{kerrisk10}, a Linux-specific I/O event notification
mechanism that polls multiple sockets with $O(1)$ complexity.  

In addition, MemEC exploits multi-threading to parallelize I/O operations.
Each entity (i.e., the coordinator, proxy, or server) uses a dedicated thread
to poll I/O events and multiple worker threads to process the received events.
It also uses multiple worker threads to send data to different destinations in
parallel (e.g., in \texttt{SET} request).  To mitigate context switching in
multi-threading, MemEC adopts an event-driven architecture.  Specifically, 
each entity consists of a multi-producer, multi-consumer event queue and a
fixed number of worker threads.  When the polling thread receives a new I/O
event, it dispatches the event, via the event queue, to one of multiple worker
threads for processing.  Each worker thread may also insert one or multiple
events to the event queue for further processing (e.g., sealing a data chunk
after a \texttt{SET} request). As multiple worker threads process the I/O
events in parallel, the overall latency of request handling is significantly
reduced.



\section{Evaluation}
\label{sec:evaluation}

We conduct extensive testbed experiments on MemEC under commodity settings.
We run MemEC on a cluster of 21 machines, including one coordinator, four
proxies, and 16 servers.  Each machine runs Ubuntu 12.04~LTS with the Linux
kernel version 3.5.0-54-generic or higher.  All servers are equipped with
Intel Core i5-3570 3.4GHz CPU, while the coordinator and all proxies are
equipped with Intel Core i5-2400 3.1GHz CPU.  Each server pre-allocates 8GB
memory to form a KV store of 128GB capacity in total.  All machines are
interconnected via a Gigabit full-duplex switch.

We use workloads derived from YCSB version~0.7.0 \cite{cooper10} to evaluate
MemEC in both normal and degraded modes.
We focus on the read-heavy and update-heavy workloads, namely A, B, C, D, and
F, as shown in Table~\ref{table:ycsb}, while we omit Workload~E since MemEC
does not support range queries.  We co-locate both YCSB clients and MemEC
proxies in the same machine. Each of the YCSB clients connects to its
co-located MemEC proxy via the loopback interface.  Before running each
workload, we load MemEC
with 10M objects using \texttt{SET} requests (called the {\em load phase}).
Each of the four clients then issues 5M requests via YCSB (i.e., 20M requests
in total), while the access pattern follows a heavy-tailed Zipf distribution
with the shape parameter 0.99.

Our experiments focus on the workloads with small objects.  We divide the set
of objects into two halves, whose value sizes are set to 8~bytes and 32~bytes.
Two clients issue requests with 8-byte values, and the other two clients issue
requests with 32-byte values.   In Experiment~3, we also consider larger value
sizes.  Also, we set the key size as 24~bytes for all objects, since the
minimum configurable key size in YCSB is 23~bytes and we use one extra byte to
distinguish between two value sizes.  For MemEC, we fix $c\!=\!16$ stripe
lists (\S\ref{subsec:decentralized}) and the chunk size as 4KB.

\begin{table}[t]
	\small
	\centering
	\begin{tabular}{|p{0.17\textwidth}|p{0.25\textwidth}|}
		\hline
		\textbf{Workload}     & \textbf{Proportions of requests}          \\ \hline
		A (Update heavy)      & 50\% \texttt{GET}, 50\% \texttt{UPDATE}   \\ \hline
		B (Read mostly)       & 95\% \texttt{GET}, 5\% \texttt{UPDATE}    \\ \hline
		C (Read only)         & 100\% \texttt{GET}                        \\ \hline
		D (Read latest)       & 95\% \texttt{GET}, 5\% \texttt{SET}       \\ \hline
		F (Read-modify-write) & 50\% \texttt{GET}, 50\% read-modify-write (\texttt{GET} and \texttt{UPDATE})  \\ \hline
	\end{tabular}
	\vspace{-6pt}
	\caption{YCSB workloads used in our experiments.}
	\vspace{-4pt}
	\label{table:ycsb}
\end{table}

To fully exploit multi-threading, we tune the numbers of threads in YCSB and
MemEC for maximum possible performance. Finally, we fix 64~threads in YCSB for
workload generation and 12~worker threads in each MemEC node (i.e., the
coordinator and each of the proxies and servers).


We plot the average results over 10 runs with 95\% confidence intervals based
on the student's $t$-distribution. Some intervals may be invisible due to
small variations.

\subsection{Performance in Normal Mode}

We start with evaluating MemEC in normal mode when no failure happens.

\paragraph{Experiment~1 (Comparisons with existing systems):} We compare MemEC
with two existing in-memory KV stores: Redis~3.0.7 \cite{redis} and
Memcached~1.4.25 \cite{memcached}.  Our objective here is {\em not} to show
which system is better than others, since all systems are implemented with
different functionalities; in fact, Redis and Memcached support more features
(e.g., object eviction) than MemEC.  Instead, our objective is to validate the
implementation of our MemEC prototype, such that its performance matches that
of state-of-the-arts.

We use the default settings for both Redis and Memcached, except that
we configure 12 worker threads in Memcached as in MemEC.  Memcached does not
support fault tolerance, while Redis uses replication for fault tolerance.
We disable the replication in Redis to maximize its performance.  For MemEC,
we disable erasure coding and set $n\!=\!10$, such that all stripe lists
contain data servers only.

\begin{figure}[t]
	\centering
	\begin{subfigure}[t]{.48\linewidth}
	\includegraphics[width=\textwidth]{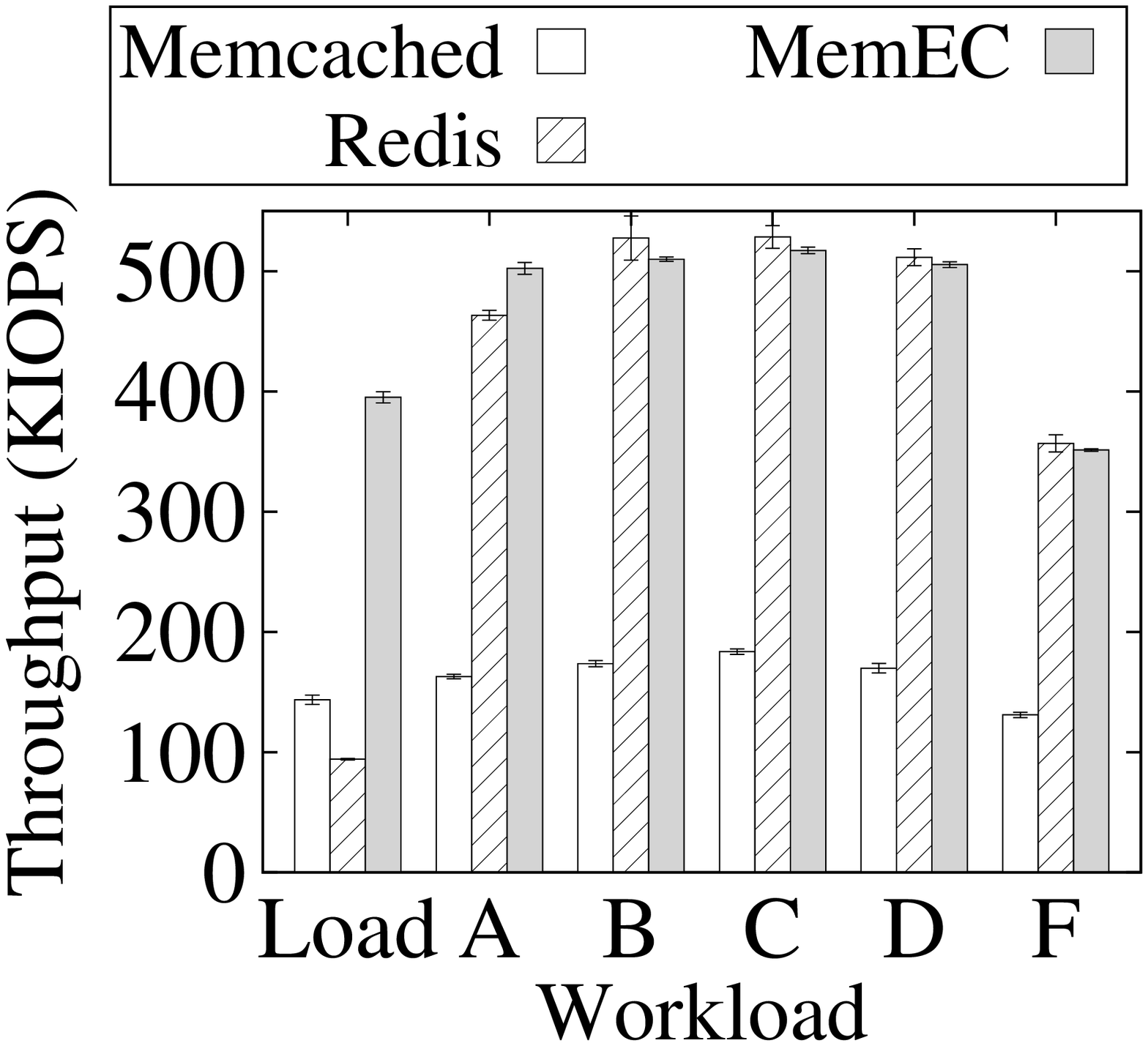}
	\caption{Aggregate throughput}
	\label{fig:norm_perf_throughput}
	\end{subfigure}
	\begin{subfigure}[t]{.48\linewidth}
	\includegraphics[width=\textwidth]{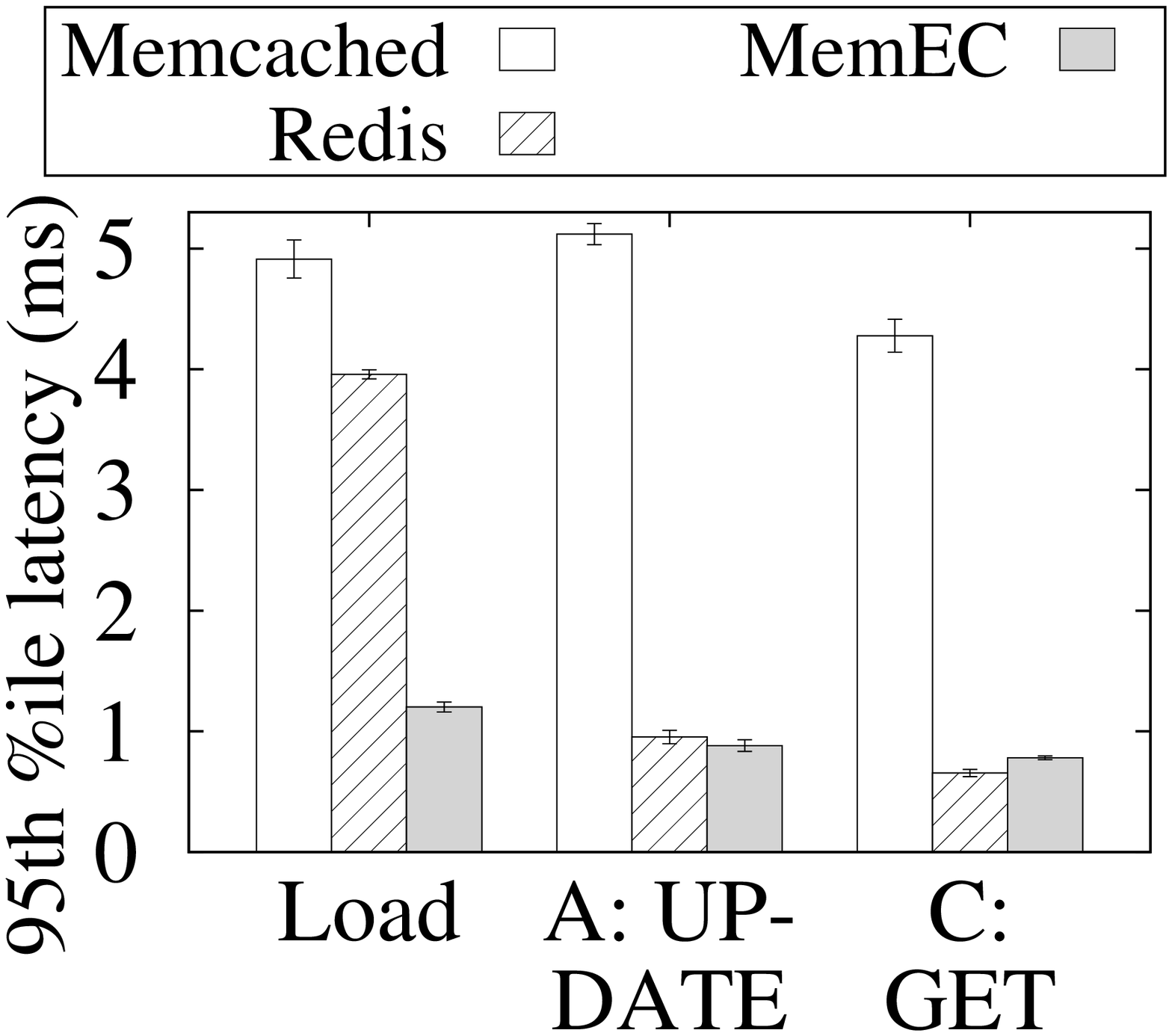}
	\caption{95th percentile latencies}
	\label{fig:norm_perf_latency}
	\end{subfigure}
	\vspace{-3pt}
	\caption{Experiment~1: Comparisons of MemEC, Redis, and Memcached.}
	\vspace{-8pt}
	\label{fig:norm_perf}
\end{figure}

Figure~\ref{fig:norm_perf_throughput} shows the aggregate throughput over all
objects for different systems.  MemEC's throughput is
2.7--3.1$\times$ of Memcached's in all cases.  The low throughput of Memcached
may be explained by its poor scalability on multi-core CPUs \cite{gunther10}.
Compared to Redis, MemEC's throughput is 4.2$\times$ of Redis's in the load
phase.  MemEC is much faster since it leverages multi-threading to handle
\texttt{SET} requests on multiple servers in parallel, while Redis
uses a {\em mostly} single-threaded design to serve requests (some slow I/O
requests are served via multi-threading) \cite{redis_latency}.  On the other
hand, MemEC's throughput is close to Redis's in Workloads~A--F: MemEC is faster
in Workload~A (which is update-heavy) by 8.4\%, while Redis is slightly
faster in Workloads~B--F (which are read-heavy) by 1.2--3.3\%.  Since reading
an object only involves a single server, there is no significant performance
difference between single- and multi-threading.

Figure~\ref{fig:norm_perf_latency} depicts the 95th percentile latencies of
different systems. In the interest of space, we only consider three specific
types of requests: \texttt{SET} in the load phase, \texttt{UPDATE} in
Workload~A, and \texttt{GET} in Workload~C.  We find that both 8-byte and
32-byte values have very similar latency results.  Thus, we directly report
the average latency over all objects (same for other experiments).
Memcached has the highest latency in all cases.  MemEC achieves
70\% lower latency than that of Redis for \texttt{SET}, while both MemEC and
Redis have comparable latencies for \texttt{UPDATE} and \texttt{GET}.

We point out that MemEC has comparable performance to 
existing in-memory storage systems proposed in the literature (e.g.,
\cite{hong13,cheng15,li15}).  For example, in a commodity cluster over a
Gigabit network, their access latencies are reportedly in the range of few
milliseconds. 

\paragraph{Experiment 2 (Erasure coding in MemEC):}
We study the erasure coding overhead in MemEC.  Here, we consider RDP
codes \cite{corbett04} and Reed-Solomon (RS) codes \cite{reed60}.  RDP codes
are double-fault tolerant.  They perform only bitwise XOR-based coding
operations and are computationally efficient.  RS codes perform
Galois Field arithmetic (\S\ref{sec:background}) and are generally more
computationally expensive than RDP codes, but they can tolerate a general
number of server failures.  We set $(n,k)\!=\!(10,8)$.  We also evaluate MemEC
when no erasure coding is used (as in Experiment~1), referred to as \textit{No
coding}, as well as Redis with 3-way replication.

\begin{figure}[t]
	\centering
	\begin{subfigure}[t]{.48\linewidth}
		\includegraphics[width=\textwidth]{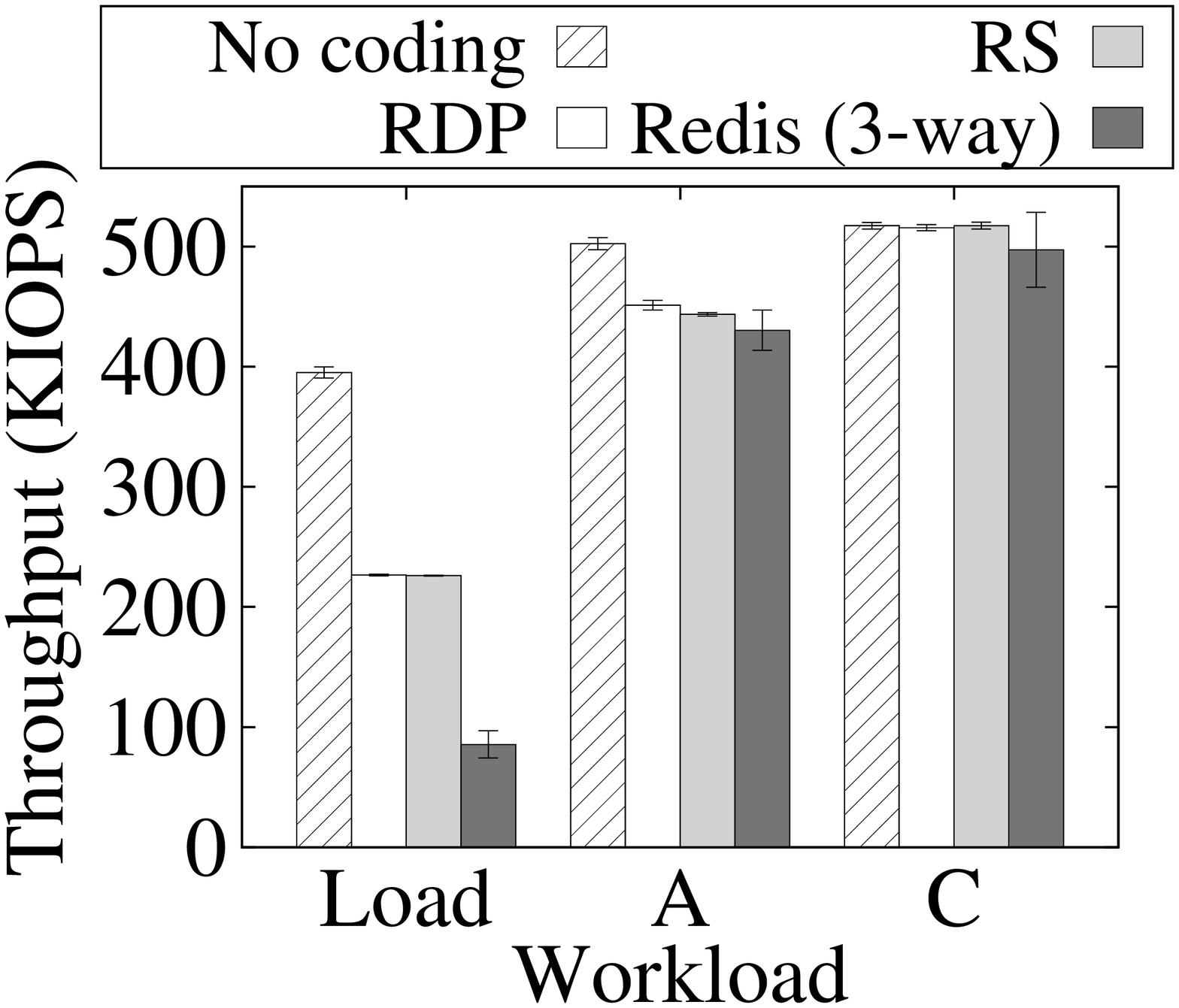}
		\caption{Aggregate throughput}
		\label{fig:encoding_throughput}
	\end{subfigure}
	\begin{subfigure}[t]{.48\linewidth}
		\includegraphics[width=\textwidth]{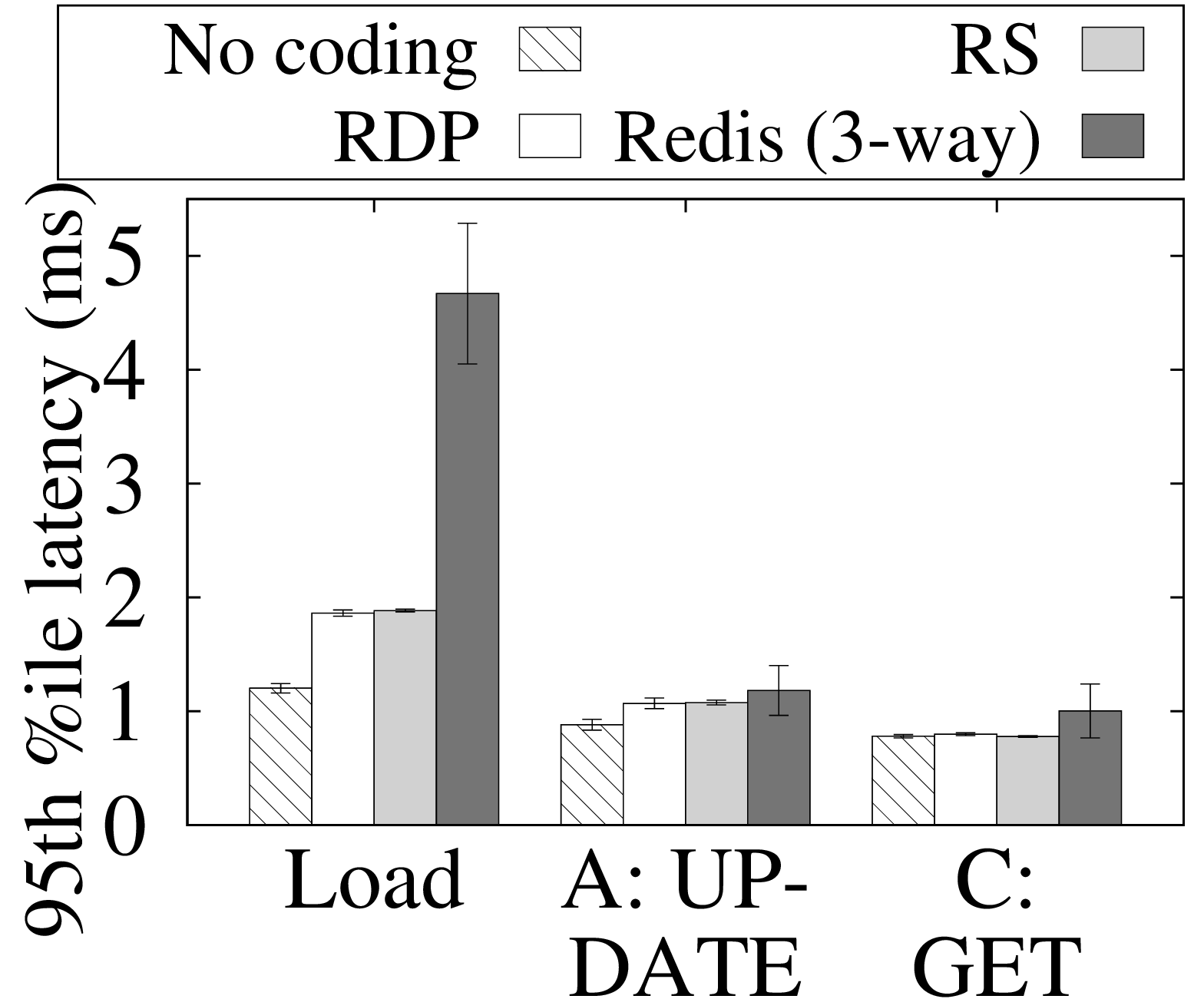}
		\caption{95th percentile latencies}
		\label{fig:encoding_latency}
	\end{subfigure}
	\vspace{-3pt}
	\caption{Experiment 2: Performance of different erasure coding schemes in
MemEC and 3-way replication in Redis.}
	\vspace{-12pt}
\end{figure}

Figure~\ref{fig:encoding_throughput} shows the aggregate throughput over all
objects for different coding schemes. We focus on the load phase, Workload~A,
and Workload~C.  In the load phase, the throughput of RDP and RS drops to
57.3\% and 57.2\% of that without coding, respectively, mainly due to the
extra traffic generated to parity servers in \texttt{SET} requests.  Also, in
Workload~A, the throughput values of RDP and RS codes are 90.0\% and 88.2\% of
that without coding, respectively.  Thus, erasure coding does not
significantly reduce the performance for \texttt{UPDATE} requests.
Furthermore, in Workload~C, both coding schemes achieve similar throughput to
that without coding, as \texttt{GET} requests only access data servers.
Overall, both RDP and RS codes have very similar results, so the coding
schemes do not significantly change the performance.

Compared to 3-way replication in Redis, we observe that the throughput of
MemEC with RDP and RS is at least 164\%, 4.0\%, and 3.9\% higher in
the load phase, Workload~A, and Workload~C, respectively.  Our results are
consistent with those in Experiment~1.

Figure~\ref{fig:encoding_latency} shows the 95th percentile latencies.
For \texttt{SET} requests, both RDP and RS codes have 55.8\% higher latencies
than without coding; for \texttt{UPDATE} requests, they are 21.7\% higher than
without coding; for \texttt{GET}, the latencies are almost identical across
coding schemes.


\begin{figure}[t]
	\centering
	\includegraphics[width=.3\textwidth]{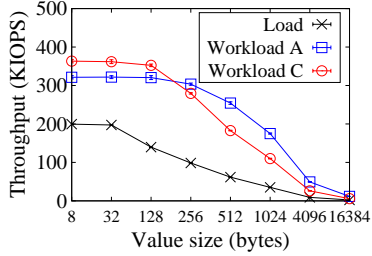}
	\vspace{-3pt}
	\caption{Experiment~3: Aggregate throughput of the load phase, Workload~A,
		and Workload~C versus value size.}
	\vspace{-8pt}
	\label{fig:value_size_throughput}
\end{figure}

\paragraph{Experiment~3 (Workloads of larger value sizes):}  We study how
MemEC performs for larger value sizes, even though it is designed for small
objects.  Specifically, in each test, we configure each of the
four clients to send objects of the same value size, which we vary from
8~bytes to 16KB across different tests.  Each client first loads 250K objects
(i.e., 1M objects in total).  Each of the four clients then issues 500K
requests (i.e., 2M requests in total) in both Workload~A and Workload~C.
Recall that if an object has size larger than the 4KB chunk size, it will be
split into fragments before being stored (\S\ref{subsec:our_data_model}).  We
use RDP with $(n,k)\!=\!(10,8)$ for erasure coding.



Figure~\ref{fig:value_size_throughput} shows the aggregate throughput versus
different value sizes.  When the value size is below 1KB, the performance is
bottlenecked by the processing overhead of small objects.  As the value size
is at least 1KB, the throughput, in terms of the amount of data being
processed per second, is saturated at 36.5MB/s, 181MB/s, and 89.0MB/s for
the load phase, Workload~A, and Workload~C, respectively.  Note that the
network bandwidth is not fully utilized for large objects.  For example, for
the load phase, we expect that four proxies can theoretically achieve an
effective network speed of at most $\frac{4\times 125}{3}\!=\!167$MB/s
(recall that a proxy first writes data via \texttt{SET} with $(n-k+1)$-way
replication (\S\ref{subsec:basic})).  We plan to optimize the implementation
for large objects as future work.

\subsection{Performance in Degraded Mode}

We now evaluate MemEC in degraded mode and examine the impact
of transient failures.  We use RDP \cite{corbett04} with $(n,k)\!=\!(10,8)$
for erasure coding.  We simulate a transient failure as network
congestion, by injecting an artificial link delay to the network interface of
a failed server for each outgoing packet via the Linux traffic controlling
tool \texttt{tc} \cite{tc}.  We configure the delay to follow a normal
distribution with mean~2ms and standard deviation~1ms.  Note that the actual
delay for determining a transient server failure varies depending on the
deployment environment, and we do not explore this issue in this work.

\begin{figure}[t]
\centering
\begin{subfigure}[t]{0.225\textwidth}
	\includegraphics[width=\textwidth]{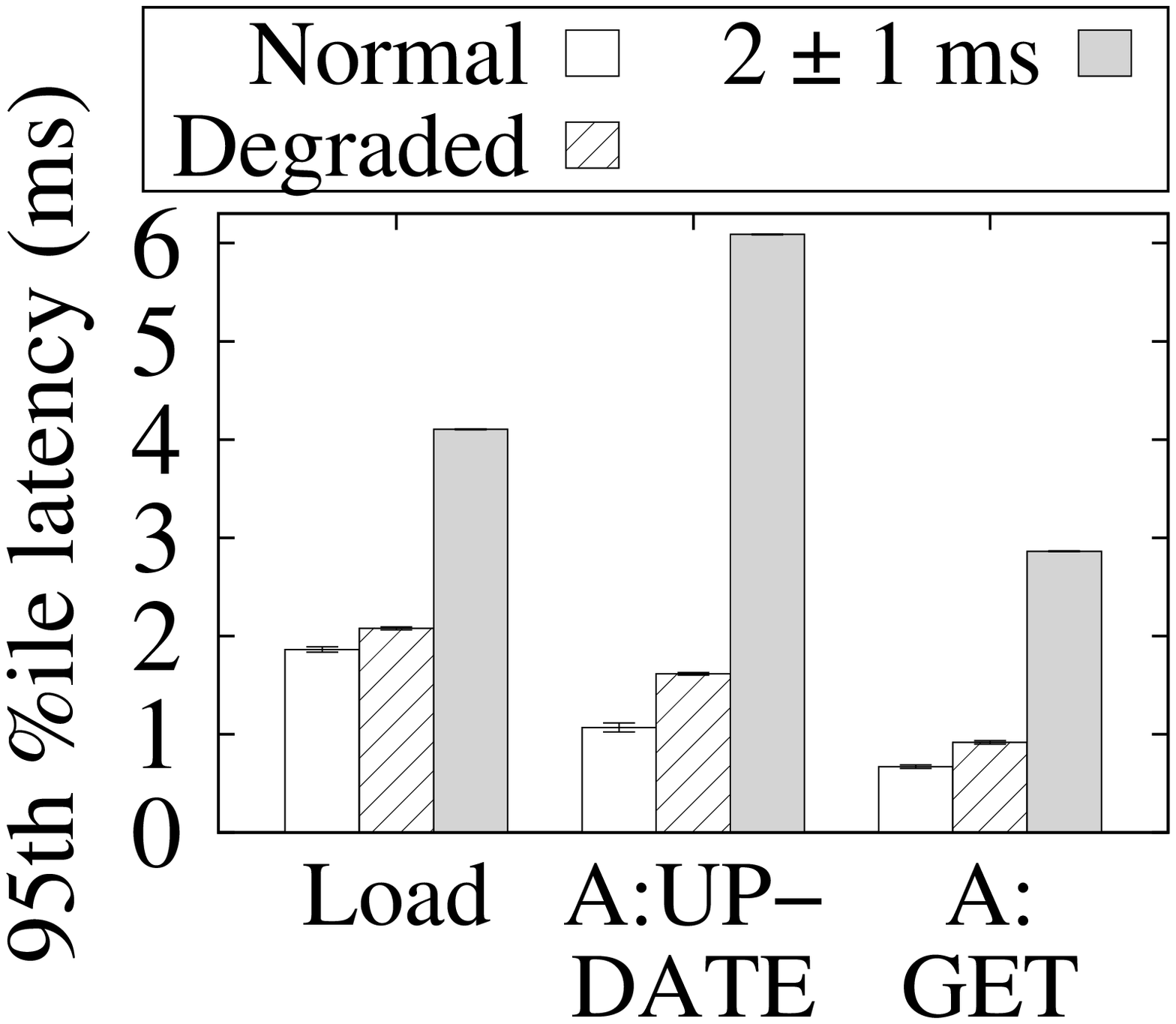}
	\caption{Before writes}
	\label{fig:degraded_set}
\end{subfigure}
~
\begin{subfigure}[t]{0.225\textwidth}
	\includegraphics[width=\textwidth]{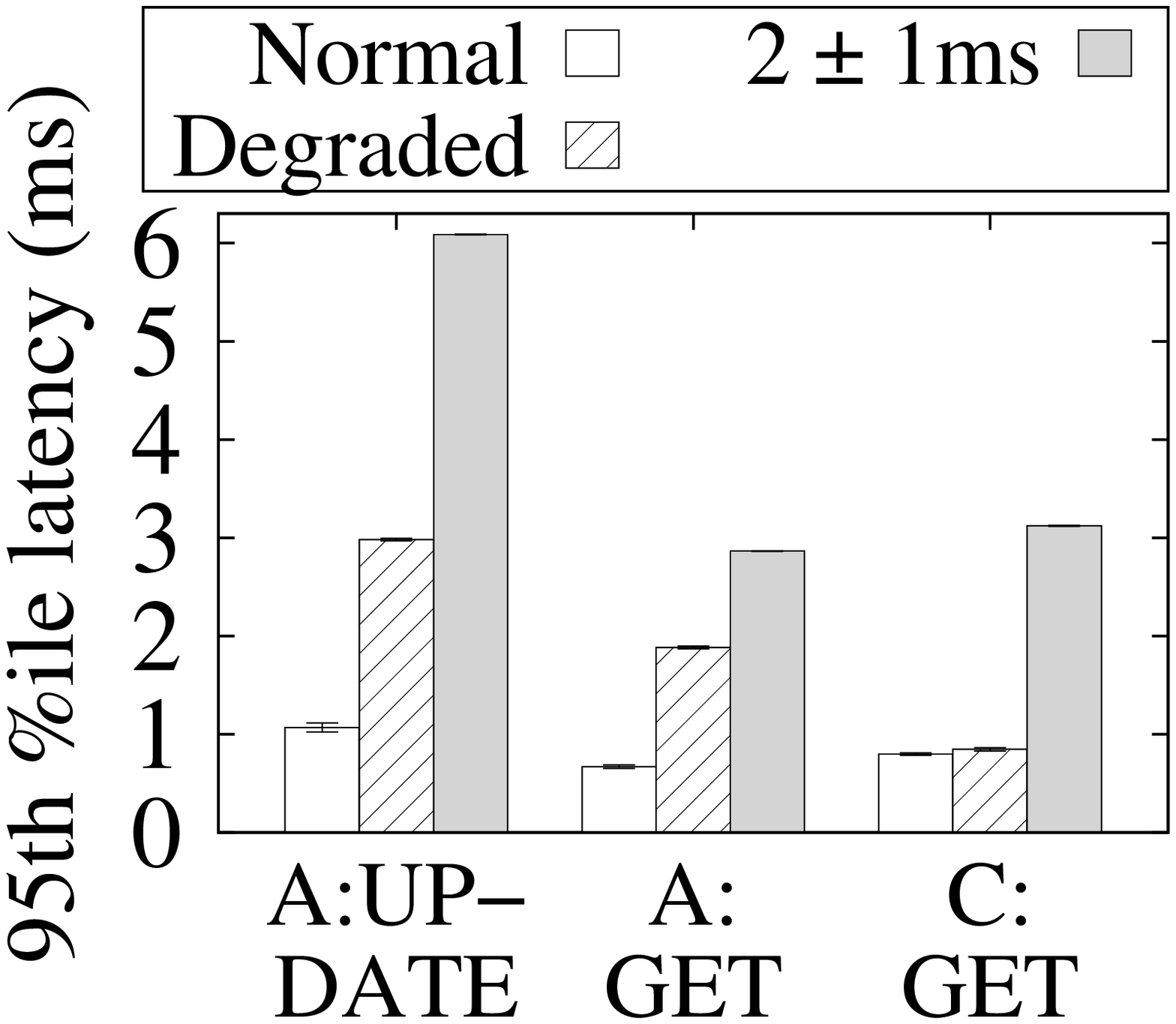}
	\caption{After writes}
	\label{fig:degraded_get_update}
\end{subfigure}
\vspace{-6pt}
\caption{Experiment~4: Impact of transient failures.}
\vspace{-6pt}
\label{fig:transient_failures}
\end{figure}

\paragraph{Experiment~4 (Impact of transient failures):}  We evaluate the
degraded requests of MemEC in the presence of transient failures.  We only
consider a single-server failure, while the results are similar for a
double-server failure.  We consider two independent cases:
\begin{itemize}[leftmargin=*] \itemsep=0pt \parskip=0pt
\item
\textbf{Before writes:}  A transient failure happens before MemEC stores any
data. We then run the load phase followed by Workload~A.  Each proxy issues
degraded \texttt{SET} requests that redirect writes from the failed server to
a redirected server in the load phase. Subsequent \texttt{GET} and
\texttt{UPDATE} requests are also redirected in Workload~A.
\item \textbf{After writes:}  We run the load phase and trigger a transient
failure after the load phase is completed.  We then run Workloads~A and C
to evaluate degraded \texttt{GET} and degraded \texttt{UPDATE}.
\end{itemize}

We compare the performance of MemEC for three settings:
(1) there is no failure (labeled as {\em Normal});
(2) there is a failure and MemEC enables degraded request handling (labeled
as {\em Degraded}); and
(3) there is a failure but degraded request handling is disabled (i.e.,
a request contacts a failed server that is under network congestion and waits
for its delayed response).

Figure~\ref{fig:degraded_set} shows the 95th percentile latencies of
\texttt{SET}, \texttt{UPDATE}, and \texttt{GET} requests when transient
failures occur before writes.  Compared to normal mode, the latency of
\texttt{SET} in the load phase increases by 11.6\% with degraded
request handing enabled.  For Workload~A, the latencies of \texttt{UPDATE} and
\texttt{GET} increase by 50.9\% and 36.9\%, respectively.
However, with degraded request handling disabled, the latencies are much
higher; for example, they increase by 469\% and 326\% in
\texttt{UPDATE} and \texttt{GET} over normal mode, respectively.

Figure~\ref{fig:degraded_get_update} shows the 95th percentile latencies of
\texttt{GET} requests in Workloads~A~and~C, and \texttt{UPDATE} requests in
Workload~A when transient failures occur after writes.  The latencies of
\texttt{GET} and \texttt{UPDATE} increase by at most 180\% and 178\%,
respectively. Note that the latencies of degraded \texttt{GET} in
Workloads~A~and~C differ, which can be explained by the need of maintaining
consistency within the same stripe. Specifically, if a redirected server
receives an \texttt{UPDATE} before \texttt{GET}, it reconstructs all failed
chunks of the same stripe before
modifying the data and parity chunks so as to avoid inconsistency caused by
concurrent access to parity chunks by \texttt{UPDATE} and reconstruction of
chunks by \texttt{GET}.  The following \texttt{GET} requests to the failed
data chunks are blocked until the reconstructed chunks are available on the
redirected server.  Thus a higher latency is observed in Workload~A, but not
in Workload~C, which involves no \texttt{UPDATE}.

Note that the impact of reconstructing failed data on a read-heavy workload is
very small.  For example, the latency of \texttt{GET} in Workload~C only
increases by 6.0\% when compared to normal mode.  MemEC performs
reconstruction at the granularity of chunks, and the reconstruction overhead
can be amortized over subsequent \texttt{GET} requests to the same
reconstructed chunks (\S\ref{subsec:fault_degraded}).

\paragraph{Experiment~5 (State transition overhead):}  We further study the
state transition overhead in the presence of a transient failure.  We consider
the transitions when a failed server switches from the normal to degraded
states (denoted as $T_{N{\to}D}$) and later when it switches from the degraded
to normal states (denoted as $T_{D{\to}N}$).  In addition to a single-server
failure, we also consider the case of a double-server failure, in which two
servers fail simultaneously.



We compare the cases with and without ongoing requests during the transitions.
For the former, we run Workload~A to generate ongoing requests until the
latency becomes stable, and trigger a transient failure.  We then restore the
failure 5s later.  We measure both the elapsed times for the state transitions
and the average latencies of \texttt{UPDATE} and \texttt{GET} requests over
the workload.

\begin{table}[t]
\small
\centering
\begin{tabular}{|c|c|c|c|}
	\hline
	\multicolumn{2}{|c|}{\multirow{2}{*}{\textbf{State transition}}}  & \multicolumn{2}{c|}{\textbf{Elapsed time (ms)}}   \\ \cline{3-4}
	\multicolumn{2}{|c|}{}                           & \textbf{Single failure}
	& \textbf{Double failure}                           \\ \hline
	\multirow{2}{*}{{$T_{N{\to}D}$}}     & With req. &  4.77 $\pm$ 0.79       &  9.24 $\pm$ 0.78                          \\ \cline{2-4}
										 & No req.   &  1.74 $\pm$ 0.09       &  4.91 $\pm$ 0.89                          \\ \hline
	\multirow{2}{*}{$T_{D{\to}N}$}       & With req. &  628.5 $\pm$ 43.9      &  667.5 $\pm$ 27.2                         \\ \cline{2-4}
										 & No req.   &  0.91 $\pm$ 0.46       &  1.10 $\pm$ 0.19                          \\ \hline
\end{tabular}
\vspace{-6pt}
\caption{Experiment~5: Average elapsed times of state transitions with 95\%
confidence intervals.}
\vspace{-12pt}
\label{table:state_transit}
\end{table}


Table~\ref{table:state_transit} first shows the elapsed times of state
transitions, averaged over 10~runs with 95\% confidence intervals.
For a single-server failure, $T_{N{\to}D}$ takes 4.77ms and
1.74ms on average with and without requests, respectively. The difference
between the two elapsed times is mainly caused by reverting parity updates of
incomplete requests.  The average elapsed times for $T_{D{\to}N}$ are 628.5ms
and 0.91ms with and without requests, respectively.  The elapsed time for
$T_{D{\to}N}$ with ongoing requests includes data migration from the
redirected server to the restored server, so it is significantly larger than
that without any request.
For a double-server failure, the transitions $T_{N{\to}D}$ and $T_{D{\to}N}$
with ongoing requests spend 9.24ms and 667.5ms, respectively.
Note that the elapsed times for $T_{N{\to}D}$ are higher than that for
$T_{D{\to}N}$ even though when there is no request.  The reason is that the
atomic broadcast for the transition $T_{N{\to}D}$ involves each failed server,
which experiences network congestion.
In all cases, a state transition incurs less than 1s, so we expect that its
overhead is limited.

\section{Related Work}
\label{sec:related}

Erasure coding has been studied in large-scale KV stores for low-redundancy
fault tolerance, such as distributed hash tables (DHTs)
\cite{dabek04,rodrigues05,chun06} and cloud storage
\cite{ceph,anderson10,lai15}.  
However, the above systems store large-size objects and perform erasure coding
on a per-object basis, which is inadequate for in-memory KV stores in which
small objects dominate as we show in this paper.

Some in-memory KV stores combine erasure coding and replication, such as LH*RS
\cite{litwin05,cieslicki06,cieslicki10} and Cocytus \cite{zhang16}.  They
store keys and metadata by replication, while storing values of multiple
objects by erasure coding.  
MemEC differs from LH*RS and Cocytus by applying erasure coding across
objects in entirety to further reduce redundancy based on the all-encoding
data model.  We also design MemEC to explicitly support efficient workflows of
degraded requests and transitions between normal and degraded modes. 




Ensuring parity consistency is critical in any erasure-coded storage
system.
Myriad \cite{chang02} leverages two-phase commit to consistently update both
data and parity chunks. 
LH*RS \cite{litwin05} uses a variant of two-phase commit to retain data delta
backups in a temporary buffer for possible rollbacks.
Aguilera et al. \cite{aguilera05} propose to keep a recent list of the
previous data update requests to resolve consistency of multiple chunks.
Pahoehoe \cite{anderson10} associates each version of object with a globally
unique timestamp to guarantee eventual consistency.
Cocytus \cite{zhang16} piggybacks the commit message of each previous request
in the current parity update and avoids the two rounds of message exchanges in
two-phase commit. 
MemEC keeps data delta backups as in LH*RS \cite{litwin05} for reverting any
inconsistent change when a server fails.

\section{Conclusion}
\label{sec:conclusion}

This paper presents MemEC, which makes a case of how to efficiently apply
erasure coding to small objects in in-memory KV storage.  We propose an
all-encoding data model that effectively reduces redundancy for fault
tolerance by up to 60\% over existing approaches.  We design and implement
MemEC to efficiently operate in both normal and degraded modes.  We evaluate
our MemEC prototype running on commodity hardware, and demonstrate its
efficiency in both normal and degraded modes under YCSB benchmarks.


\paragraph{Acknowledgment:} We would like to thank our shepherd, Nuno
Pregui\c{c}a, and the anonymous reviewers for their valuable comments.  This
work was supported in part by grants from Huawei Technologies (YB2014110015)
and ITS/113/14 from the ITF of HKSAR.

{\small
\bibliographystyle{abbrv}
\bibliography{paper}
}

\end{document}